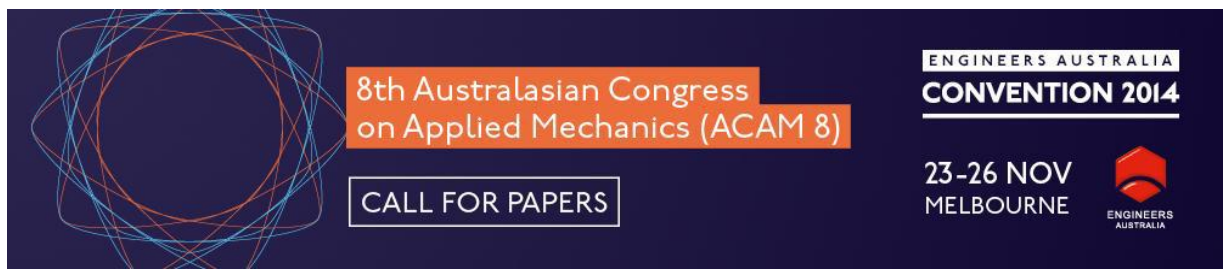

# Numerical modelling of sandstone uniaxial compression test using a mix-mode cohesive fracture model

**Yilin Gui\*, Jayantha Kodikara and Ha H. Bui**

*Department of Civil Engineering, Monash University, Clayton, VIC, Australia*

*\*Corresponding author. Email:* yilin.gui@monash.edu

*Abstract*: A mix-mode cohesive fracture model considering tension, compression and shear material behaviour is presented, which has wide applications to geotechnical problems. The model considers both elastic and inelastic displacements. Inelastic displacement comprises fracture and plastic displacements. The norm of inelastic displacement is used to control the fracture behaviour. Meantime, a failure function describing the fracture strength is proposed. Using the internal programming FISH, the cohesive fracture model is programmed into a hybrid distinct element algorithm as encoded in Universal Distinct Element Code (UDEC). The model is verified through uniaxial tension and direct shear tests. The developed model is then applied to model the behaviour of a uniaxial compression test on Gosford sandstone. The modelling results indicate that the proposed cohesive fracture model is capable of simulating combined failure behaviour applicable to rock.

## 1 Introduction

It has been proven that linear elastic fracture mechanics (LEFM) is a useful approach for addressing fracture problems if it has a crack-like flaw in the material and non-linear zone in front of a crack tip is sufficiently small to consider it is negligible [1]. In addition, LEFM also assumes that the intact material behaviour is linear elastic. However, for many geomaterials such as soils, rocks, and concrete and cement stabilised rock aggregate and soil, it may be unrealistic to consider the size of non-linear zone is negligible and intact material to be always linear elastic (e.g., [1]). To overcome some of these shortcomings, cohesive fracture model (also referred to as fictitious crack model), first proposed by Duddale [2], Barenblatt [3], has been advanced (e.g., [1], [4], [5]). In Mode-I fracture the cohesive fracture model and its constitutive behaviour may be described as shown in Figure 1. The fracture is considered as two components, i.e. real crack and fictitious crack, which is also known as the process zone. Accordingly, two crack tips, namely real and virtual crack tips, are considered as shown in Figure 1. The process zone is the zone between the real and fictitious crack tips and comprises the material that is partially damaged but is still able to transfer load [5] across the fracture. The crack opening behaviour is governed by the value of opening displacement and strength of the material. The initial hardening behaviour may be linear elastic when the opening displacement value is smaller than a critical value (i.e. $w_{c1}$ in Figure 1(b), (c), (d), (e), (f)). The crack surface traction at the critical opening is the material tensile strength (i.e. $\sigma_t$ in Figure 1(b), (c), (d), (e), (f)). For openings larger than $w_{c1}$, the bridging stress across fracture will decreases featuring softening behaviour and will become zero at a limiting displacement (i.e. $w_{c2}$ in Figure 1(b)).

Generally, the initial hardening response is relatively small in comparison to the softening response, which has, therefore, received more attention in the past (e.g., [1]). To explain the softening behaviour, several softening laws have been developed including mono-linear [5], bi-linear [5], trapezoid shaped [7], rectangular shaped [8] and exponential laws [9]. Although there are advancements in cohesive fracture model development, only few models have considered the mix-mode fracture failure in materials. For instance, Kazerani and Zhao [9] proposed a cohesive fracture model for rock; however, there is no failure function defined and no plastic deformation considered for crack opening. In this paper, a cohesive fracture model that takes into account tensile, shear and compressive behaviour combined with an evolutionary failure model is presented. The cohesive fracture model is implemented using a hybrid of finite difference-discrete element method. The verification of the implementation of the model is performed though uniaxial tension and shear tests of a rock. At last, the model is applied to simulate uniaxial compression test of Gosford sandstone.

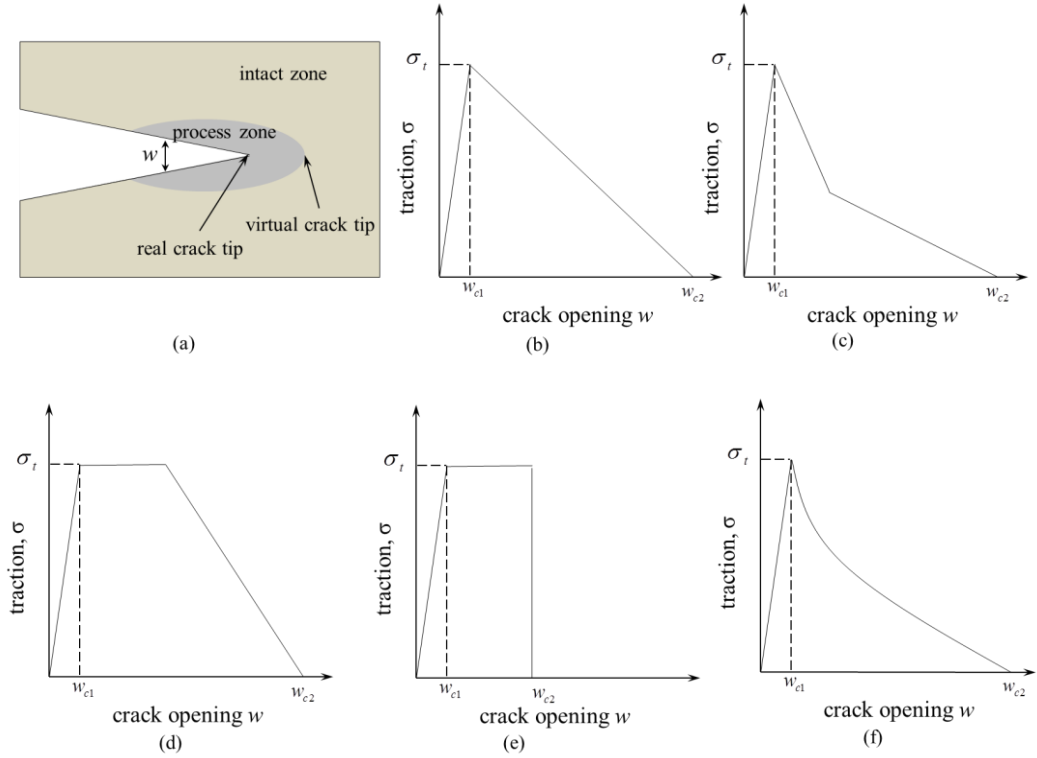

**Figure 1**: (a) General illustration of a cohesive fracture model and (b)-(f) different opening constitutive models

## 2 The cohesive fracture model

The cohesive fracture model presented herein is the extension of the framework of interface constitutive model presented in [10] that takes into account the cohesive effect on both tension and shear. The hardening stage is as assumed to be linear elastic and the emphasis is placed on the treatment of the softening stage.

The tensile loading could occur under Mode-I tension (pure tension), mixed-mode tensile/shear loading and (or) during unloading and reloading of these modes. When a material is undergoing loading, its displacement (i.e. $\mathbf{u}$) can be partitioned into elastic (i.e. $\mathbf{u}^e$) and inelastic part (i.e. $\mathbf{u}^i$) as

$$\mathbf{u} = \mathbf{u}^e + \mathbf{u}^i \tag{1}$$

The inelastic displacement can be further decomposed into plastic displacement ($\mathbf{u}^p$) which is irreversible and fracturing displacement ($\mathbf{u}^f$) as:

$$\mathbf{u}^i = \mathbf{u}^p + \mathbf{u}^f \tag{2}$$

Then the norm of the inelastic displacement ($u^{ieff}$) is defined as:

$$u^{ieff} = \|u^i\| = \|u^p + u^f\| = \sqrt{(u_n^{i\,2} + u_s^{i\,2})} \tag{3}$$

where $u_n^i$ and $u_s^i$ are the normal and shear inelastic displacement along a fracture interface.

In the tensile loading condition, the governing variables are the tensile strength ($\sigma^t$), and the norm of the inelastic displacement. The tensile strength is a linear function of the norm of the inelastic displacement which can be written as:

$$\sigma_t\left(u^{ieff}\right) = \begin{cases} \sigma_{t0}\left(1 - \dfrac{u^{ieff}}{w_\sigma}\right) & u^{ieff} < w_\sigma \\ 0 & u^{ieff} \geq w_\sigma \end{cases} \quad (4)$$

and

$$w_\sigma = \frac{2G_f^I}{\sigma_{t0}} \quad (5)$$

where $w_\sigma$ is the ultimate norm inelastic displacement corresponding to zero tensile strength, $\sigma_{t0}$ the initial tensile strength and $G_f^I$ the Mode-I fracture energy. The ultimate norm of inelastic displacement is corresponding to the threshold condition where a crack is fully developed and material is no more capable of transferring stress. The initial tensile strength is the stress at which the cohesive zone starts to develop and the crack starts to undergo softening. The tensile strength evolution during crack opening is also illustrated in Figure 2(a).

In order to cater for stiffness degradation during softening, a micro damage variable $D$ is introduced as the percentage of fracture surface (i.e. $A_f$) to overall interface area (i.e. $A_0$), and can be calculated as:

$$D = \frac{A_f}{A_0} = 1 - \frac{k_{ns}}{k_{n0}} \quad (6)$$

where $k_{ns}$ and $k_{n0}$ are the degraded and initial normal stiffness, respectively. The degraded normal stiffness ($k_{ns}$) can be computed as:

$$k_{ns} = \frac{\sigma}{u - u^p} = \frac{\sigma_t\left(u^{ieff}\right)}{u^e + u^p + u^f - u^p} = \frac{\sigma_t\left(u^{ieff}\right)}{\sigma_t\left(u^{ieff}\right)/k_{n0} + (1 - \eta)u^{ieff}} \quad (7)$$

where $\eta$ is the ratio of irreversible inelastic normal displacement to the total value of inelastic displacement (i.e. $u^p$), which can be determined experimentally using pure Mode-I test through $\eta = u^p / u^i$. The stress-displacement relationship of the interface in the normal direction is expressed as:

$$\sigma_n = k_{ns}\left(u_n - u_n^p\right) = \alpha k_{no}\left(u_n - u_n^p\right) \quad (8)$$

where $\alpha$ is the integrity parameter defining the relative active area of the fracture and it is related to the damage variable $D$ through the following relation:

$$\alpha = 1 - \frac{|\sigma_n| + \sigma_n}{2\sigma_n} D \quad (9)$$

It can be seen from Equation (9) that the activation of the micro damage variable (D) is controlled through the fraction normal stress, which is activated in tension ($\sigma_n > 0$) and deactivated in compression ($\sigma_n < 0$). Thus, both the normal and shear stiffness in tension can be degraded, while they are kept unchanged in compression.

In the similar fashion, the governing variables for the shear loading are cohesion $c$ (the contribution of normal stress to shear strength can be neglected if the friction angle is taken to be zero), and the norm of the inelastic displacement $u^{ieff}$. As the crack propagated, the cohesion is degraded and can be expressed as a linear function of the norm of the inelastic displacement $u^{ieff}$ as:

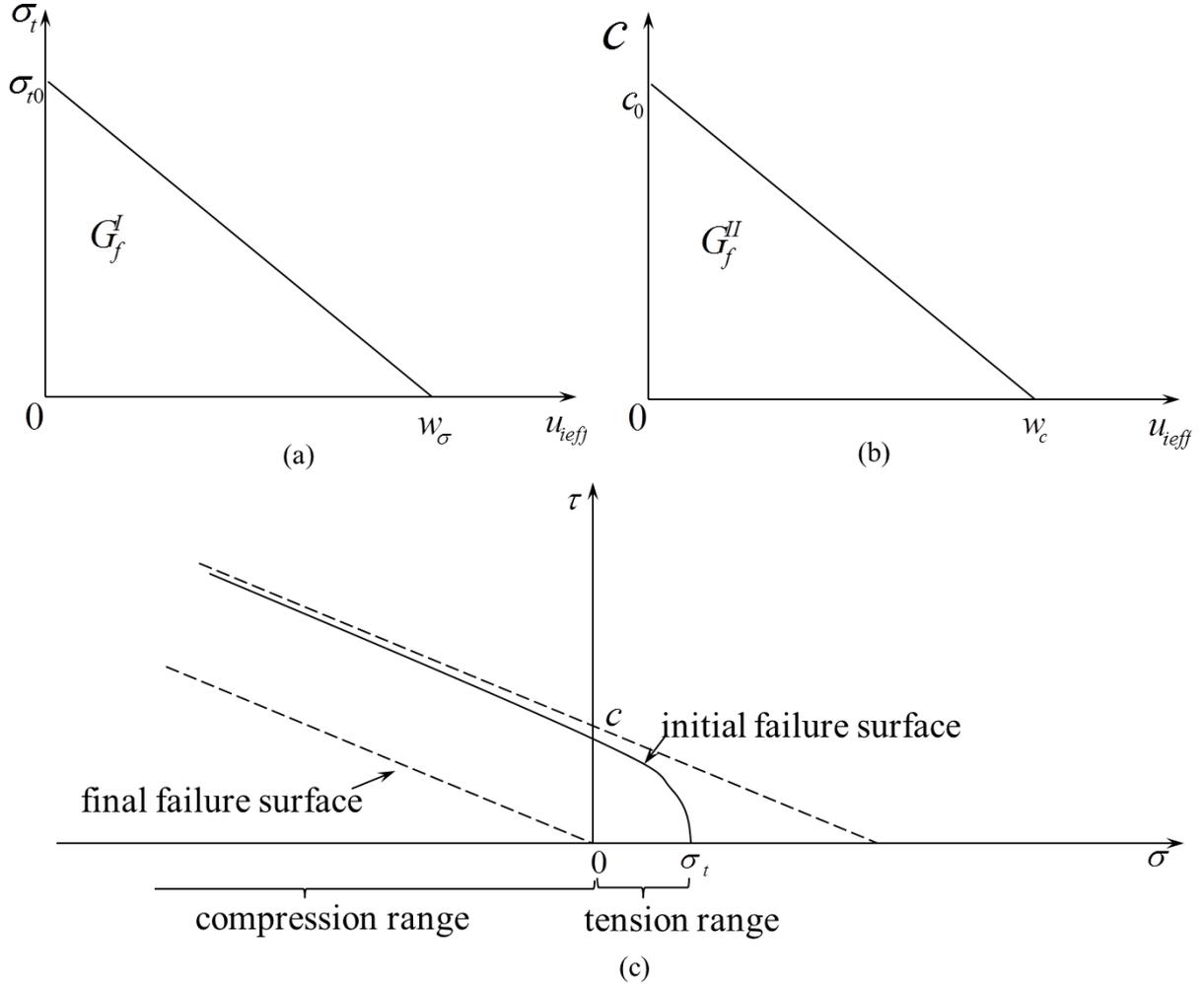

**Figure 2**: (a) Tensile and (b) compressive shear softening laws, (c) cohesive fracture failure surface

$$c(u^{ieff}) = \begin{cases} c_0\left(1 - \dfrac{u^{ieff}}{w_c}\right) & u^{ieff} < w_c \\ 0 & u^{ieff} \geq w_c \end{cases} \quad (10)$$

and

$$w_c = \dfrac{2G_f^{II}}{c_0} \quad (11)$$

where $w_c$ is the ultimate norm of inelastic displacement corresponding to zero cohesion (see Figure 2(b)), $c_0$ the initial cohesion and $G_f^{II}$ is the Mode-II fracture energy dissipated during shear at high confining normal stress (i.e. without the influence of the tensile loading regime). The ultimate norm of inelastic displacement corresponding to zero cohesion gives the threshold condition at which the shear crack is fully developed and the material is no more capable to transfer cohesion as shown in Figure 1(b). Similar to the softening treatment under tensile loading, the degraded shear stiffness ($k_{ss}$) can be written:

$$k_{ss} = \alpha k_{s0} \qquad (12)$$

And the shear stress ($\tau$) is computed by:

$$\tau = k_{ss}(u_s - u_s^p) = \alpha k_{so}(u_s - u_s^p) \qquad (13)$$

In the current cohesive fracture model, the evolution of the failure function is based on the norm of inelastic displacement which is acting as the softening parameter in the model. The failure function is defined as:

$$f = \tau^2 - 2c\tan(\varphi)(\sigma_t - \sigma) - \tan^2(\varphi)(\sigma^2 - \sigma_t^2) = 0 \qquad (14)$$

where $\varphi$ is the interface friction angle. Only the stress state hit the failure envelop defined by the failure function, will the fracture interface yield. Due to the evolution of tensile strength and cohesion, the failure surface shrinks, as shown in Figure 2(c).

## 3 Verification and application of the cohesive fracture model

The verification of the proposed cohesive fracture model is performed through uniaxial tension and direct shearing tests of sandstone.

### 3.1 Uniaxial tension

The material used for the verification is Transjurane sandstone, which was previously utilised by Kazerani et al. [11, 12]. The material properties are listed in Table 1. $\rho$ is the density (kg/m$^3$), $E$ is the elastic modulus (GPa), $\upsilon$ is the Poisson's ratio, $\varphi$ is the friction angle (degree), $d$ is the dilation angle (degree), $k_{n0}$ and $k_{s0}$ are the initial normal and shear stiffness (GPa/m), $\sigma_{t0}$ and $c_0$ are the initial tensile strength and cohesion (MPa), $w_\sigma$ and $w_c$ are the critical norm of inelastic displacement (m) corresponding to zero tensile strength and cohesion, respectively.

**Table 1:** Parameters used in uniaxial tension test

| $\rho$ | $E$ | $\upsilon$ | $\varphi$ | $d$ | $k_{n0}$ | $k_{s0}$ | $\sigma_{t0}$ | $c_0$ | $w_\sigma$ | $w_c$ |
|---|---|---|---|---|---|---|---|---|---|---|
| 2600 | 12.5 | 0.3 | 41 | 10 | 2.2321×10$^5$ | 6.573×10$^4$ | 2.8 | 8.5 | 2.8×10$^{-5}$ | 1.205×10$^{-5}$ |

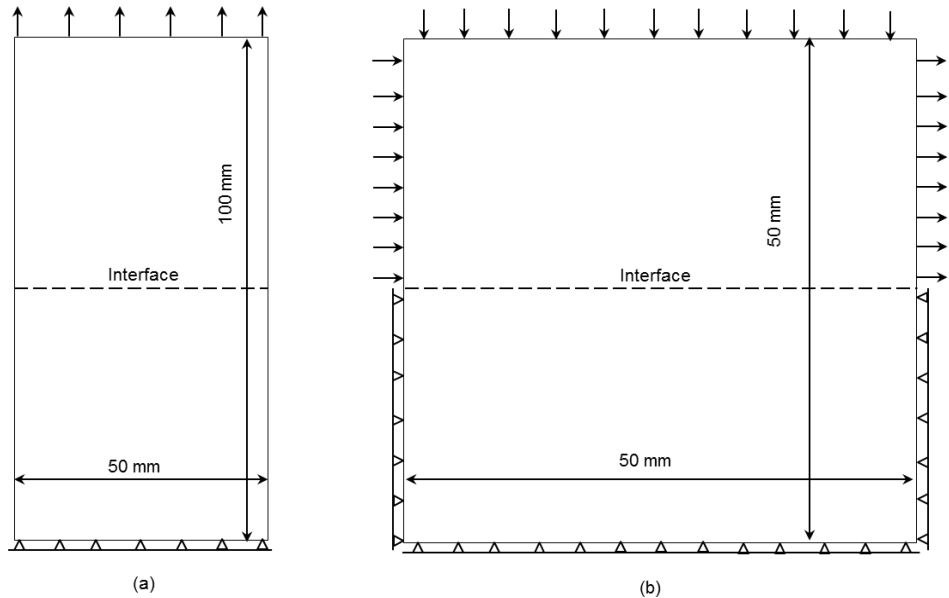

**Figure 3**: (a) uniaxial tension test, (b) direct shear test

The uniaxial tension test is performed on a rectangular sample having dimension of 100×50 mm as shown in Figure 3 (a). An interface is imbedded in the middle of the sample. In order to simulate Mode-I cracking, a tensile load is applied on the top of the sample while the bottom is fixed along both horizontal and vertical direction. The uniaxial tension is simulated using the material properties in Table 1 and the simulation results are presented in Figure 4. Figure 4(a) gives the comparison of numerical and analytical results on the opening displacement and the tensile stress on the crack. The tensile stress initially increases with opening until reaching the peak stress (2.78 MPa) which is almost equal to the initial tensile strength of the crack (as shown in Table 1). After that, the softening stage starts and the tensile stress on the crack surface decreases with opening and decreases to zero when the opening is around $2.8 \times 10^{-5}$ m. Figure 4(b) shows the evolution of the micro damage variable and the integrity parameter with the norm of inelastic displacement. The changes of micro damage variable and integrity parameter are opposite and damage variable increases from 0 to 1 but the integrity parameter decreases from 1 to 0. Figure 4(c) demonstrates the degradation of the normal stiffness with the norm of inelastic displacement. The simulated tensile strength with norm of inelastic displacement is shown Figure 4(d). As can be seen from the simulation results, all the simulation results have good agreement with analytical values.

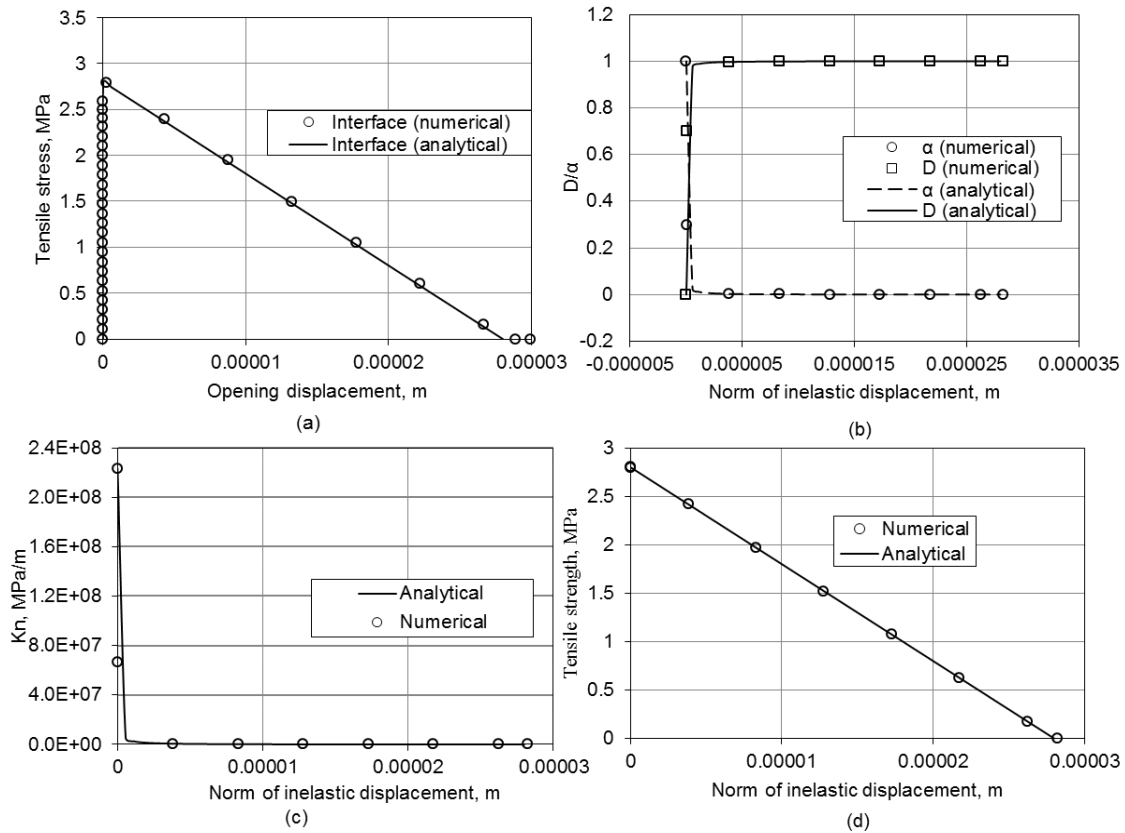

**Figure 4**: The simulation result of uniaxial tension: (a) interface tensile stress vs interface opening displacement, (b) the damage variable and $\alpha$ vs $u_{ieff}$, (c) interface normal stiffness vs $u_{ieff}$ and (d) interface tensile strength vs $u_{ieff}$

## 3.2 Shearing test

The shearing test is also conducted using Transjurane sandstone. The model geometry and boundary condition are shown in Figure 3(b). The dimension of the model is 50×50 mm. In order to keep the interface in contact during shearing, a vertical load is applied on the top of the model. Meantime, to avoid the contribution of normal stress on the fracture to shear strength, the friction angle adopted for the shearing test is zero. Figure 5 shows the simulation results conducted using the same material properties given in Table 1. Comparing to the uniaxial tension, the integrity parameter $\alpha$ in the shearing test is constant because the normal stress on the interface is compressive due to the load on the top boundary; thus, the micro damage variable *D* cannot be activated for the stiffness. Accordingly both normal and shear stiffness is kept unchanged and equal to their initial values.

### 3.3 Uniaxial compression test

The uniaxial compression test is applied on Gosford sandstone. The input parameters for the numerical model are listed in Table 2. The specimen has dimension of 100×50 mm and the boundary conditions are shown in Figure 6(a). In the simulation, the specimen is discretised into particles with size of 2 mm. Interfaces are inserted along the edges between neighbouring particles and their behaviour are in accordance with the proposed the cohesive fracture model. In the simulation, a constant velocity was applied on the top of the specimen.

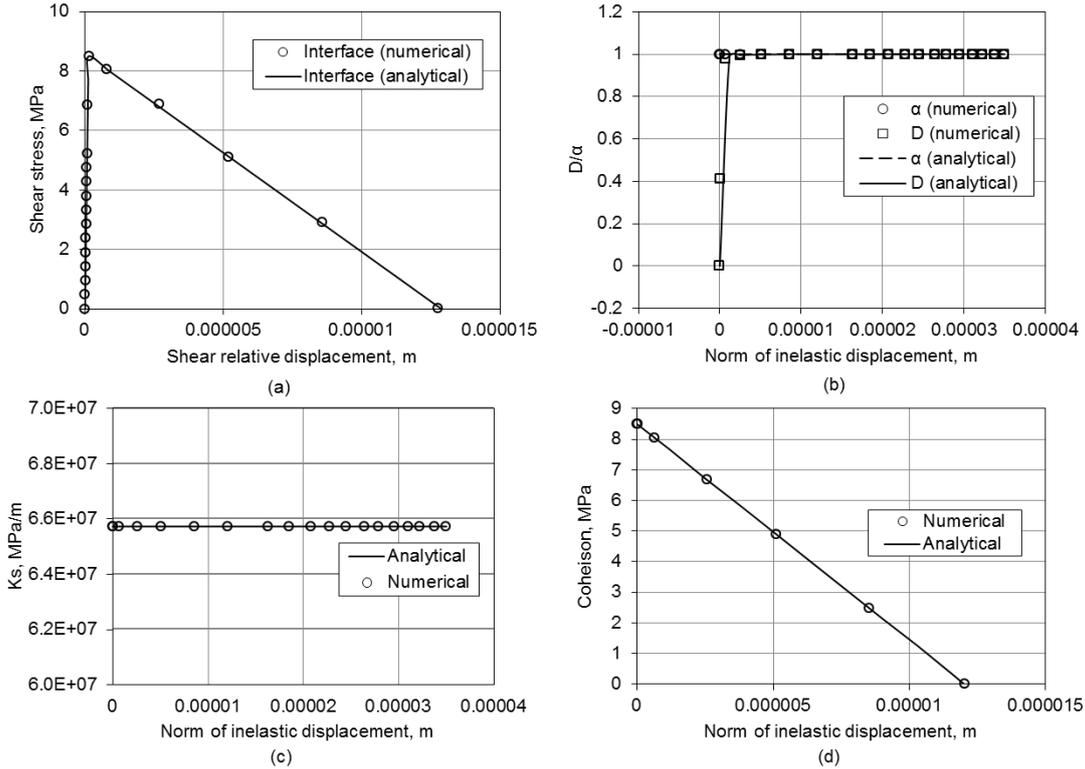

**Figure 5**: The simulation result of shearing test: (a) interface shear stress vs interface shear displacement, (b) the damage variable and *α* vs $u_{ieff}$, (c) interface shear stiffness vs $u_{ieff}$ and (d) interface cohesion vs $u_{ieff}$

**Table 2**: Parameters used in uniaxial tension test

| $\rho$ | $E$ | $\upsilon$ | $\varphi$ | $d$ | $k_{n0}$ | $k_{s0}$ | $\sigma_{t0}$ | $c_0$ | $w_\sigma$ | $w_c$ |
|---|---|---|---|---|---|---|---|---|---|---|
| 2600 | 7.0 | 0.25 | 40 | 5 | 6.0×10³ | 3.0×10³ | 6.0 | 15.0 | 1.0×10⁻⁴ | 1.5×10⁻⁴ |

The simulation result is shown in Figure 6(b). A reasonable agreement is obtained. The peak stress is same as the experimental peak stress. The discrepancy of the pre-peak curves is considered to be from the elastic constitutive model used for particles.

## 4 Conclusion

In this paper, a cohesive fracture model considering compression, tension and shear is formulated. The fracture model uses an effective inelastic displacement to relate tensile strength, cohesion, and stiffness. A micro damage variable is used to degrade both normal and shear stiffness under tension. The cohesive fracture model is successfully implemented in coupled finite difference and discrete element code (i.e. UDEC) through FISH and verified using uniaxial tension, shearing test and uniaxial compression test. A good agreement between numerical simulation and experimental results is achieved.

## Acknowledgements

The authors gratefully acknowledge the support of funding from the Monash Engineering Seed Funding Scheme 2014.

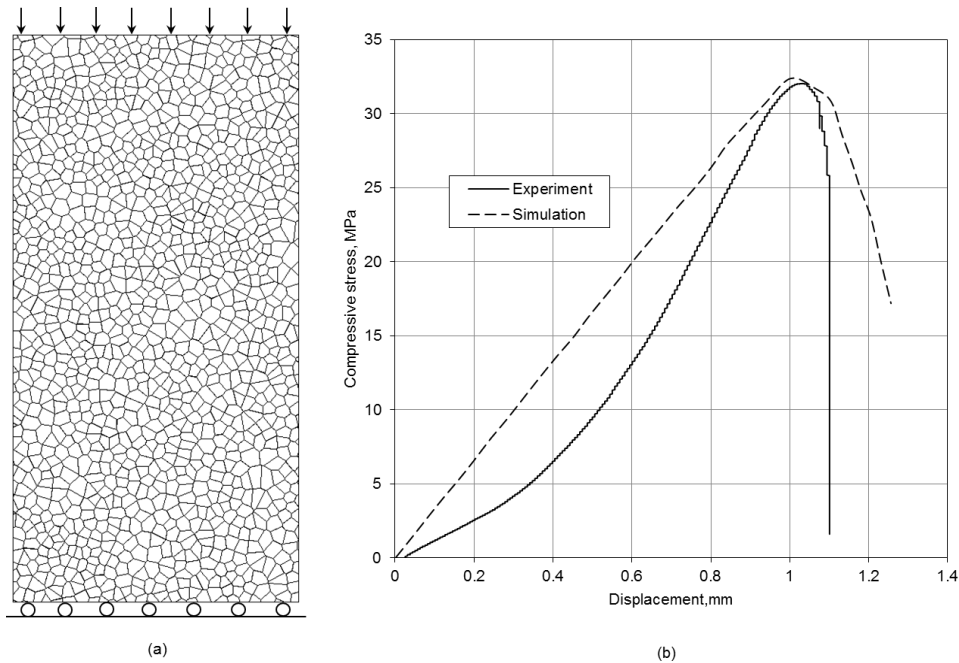

**Figure 6**: Uniaxial compression test: (a) numerical model and boundary condition of uniaxial compression test of Gosford sandstone, (b) comparison of experimental and simulation vertical displacement and compressive stress.